\documentclass[aps,prl,twocolumn,superscriptaddress,showpacs]{revtex4}  

\usepackage{graphicx}
\usepackage{epstopdf}
\usepackage{color}
\definecolor{orange}{RGB}{255,127,0}
\definecolor{blue2}{RGB}{33,114,173}
\begin{document}

\title{DMI meter: Measuring the Dzyaloshinskii-Moriya interaction inversion in Pt/Co/Ir/Pt multilayers}


\author{A.~Hrabec}
\affiliation{School of Physics and Astronomy, University of Leeds, Leeds LS2 9JT, United Kingdom}
\author{N.~A.~Porter}
\affiliation{School of Physics and Astronomy, University of Leeds, Leeds LS2 9JT, United Kingdom}
\author{A.~Wells}
\affiliation{School of Physics and Astronomy, University of Leeds, Leeds LS2 9JT, United Kingdom}
\author{M.~J.~Benitez~Romero}
\affiliation{School of Physics and Astronomy, University of Glasgow, G12 8QQ, United Kingdom }
\author{G.~Burnell}
\affiliation{School of Physics and Astronomy, University of Leeds, Leeds LS2 9JT, United Kingdom}
\author{S.~McVitie}
\affiliation{School of Physics and Astronomy, University of Glasgow, G12 8QQ, United Kingdom }
\author{D.~McGrouther}
\affiliation{School of Physics and Astronomy, University of Glasgow, G12 8QQ, United Kingdom }
\author{T.~A.~Moore}
\affiliation{School of Physics and Astronomy, University of Leeds, Leeds LS2 9JT, United Kingdom}
\author{C.~H.~Marrows}
\email[]{c.h.marrows@leeds.ac.uk}
\affiliation{School of Physics and Astronomy, University of Leeds, Leeds LS2 9JT, United Kingdom}


\date{\today}

\begin{abstract}
  We describe a field-driven domain wall creep-based method for the quantification of interfacial Dzyaloshinskii-Moriya interactions (DMI) in perpendicularly magnetized thin films. The use of only magnetic fields to drive wall motion removes the possibility of mixing with current-related effects such as spin Hall effect or Rashba field, as well as the complexity arising from lithographic patterning. We demonstrate this method on sputtered Pt/Co/Ir/Pt multilayers with a variable Ir layer thickness. By inserting an ultrathin layer of Ir at the Co/Pt interface we can reverse the sign of the effective DMI acting on the sandwiched Co layer, and therefore continuously change the domain wall (DW) structure from right- to the left-handed N\'{e}el wall. We also show that the DMI shows exquisite sensitivity to the exact details of the atomic structure at the film interfaces by comparison with a symmetric epitaxial Pt/Co/Pt multilayer.
\end{abstract}


\maketitle


The Dzyaloshinskii-Moriya interaction (DMI) \cite{dzyaloshinsky1958,moriya1960} has recently returned to prominence due to recent findings in the field of magnetic domain wall (DW) motion. Initially, DWs in Permalloy nanowires were widely studied \cite{ganieee2000,klauiprl2005,yamaguchiprl2004,hayashi2007,meier2007,lepadatu2009}, but materials with out-of-plane (OOP) anisotropy promised even higher interaction between the current and DWs \cite{boulle2008,alvarezprl2010}. It was subsequently shown that broken spatial symmetry plays an extremely important role in the current-induced DW propagation process in OOP materials \cite{Moore_APL2008,Miron_Nature2010,Buhrman_Science}. However, it has been pointed out that Bloch walls, which simple magnetostatic considerations predict to be the stable DW form in such materials \cite{malozemoff1979magnetic}, do not have the appropriate spin texture for an efficient Slonczewski-like torque \cite{khvalkovskiy2013matching}. This has been demonstrated by an application of a longitudinal magnetic field which distorts the Bloch wall towards the N\'{e}el wall structure \cite{haazen2013domain} leading to much more efficient DW motion \cite{Martinez_APL}. This is of importance in the efficient and reliable operation of technologies such as racetrack memories \cite{parkin2008}.

Soon after this demonstration, a series of theoretical \cite{Thiaville_DMI,Kim_ChiralityFromSOI} and experimental works \cite{emori2013current, ryu2013chiral,lee2014,torrejon2014} showed that a magnetic field that transforms a Bloch wall into a N\'{e}el wall can exist intrinsically due to the broken inversion symmetry at the interface. This effective field arises from the DMI as a result of high spin orbit coupling linking the broken inversion symmetry at the interface to the spin structure \cite{fert1991magnetic,crepieux1998}. In contrast to the Heisenberg interaction (usually written as $- J \mathbf{S}_1 \cdot \mathbf{S}_2$ with $J$ being the exchange integral), which favours collinear alignment of neighbouring spins $\mathbf{S}_1$ and $\mathbf{S}_2$, the DMI can be expressed as $- \mathbf{D} \cdot {\mathbf{S}_{1}} \times \mathbf{S}_2$, thus preferring an orthogonal orientation of $\mathbf{S}_1$ and $\mathbf{S}_2$ with a given chirality depending on the direction of the DM vector $\mathbf{D}$. This interaction is equivalent to a magnetic field acting across the DW and establishes a N\'{e}el wall of fixed chirality which dictates the direction of DW motion under the influence of a spin Hall torque. This interfacial effect has been experimentally demonstrated by several \textit{in situ} studies on epitaxial bilayers \cite{Kubetzka_2003,Bode,chen2013tailoring}. The DMI also plays a crucial role in bulk material systems with broken inversion symmetry producing exotic magnetization textures such as helices or skyrmions \cite{rossler2006,nagaosa2013topological}. Skyrmions have been created on the atomic scale using the interfacial DMI in monolayer of Fe on Ir \cite{heinze2011spontaneous}. It has been predicted that skyrmions have a great potential for applications as magnetic memories due to their size and extremely low operational electric currents \cite{sampaio2013nucleation,iwasaki2013current}. Therefore finding the means for \textit{ex situ} studies of materials with high and tunable DMI is of a high interest.

Here we report a simple magnetic field-based method for DMI quantification in thin films with OOP magnetic anisotropy, and demonstrate its use by measuring the DMI inversion in Pt/Co/Ir/Pt multilayers with variable Ir thickness. Since crystallographically ordered Pt/Ni and Ir/Ni bilayers exhibit DMIs of opposite sign \cite{chen2013tailoring}, the effective DMI in the Co layer can be potentially enhanced by placing the Pt and Ir layers on either side, i.e. using two DMI-active layers. Avoiding the use of currents to drive DW motion makes the method simple to implement, since it can be applied to sheet films and lithography is not required. Moreover, it makes the interpretation of the data much more straightforward, since the complexity of the interplay of spin-transfer, Rashba, and spin Hall torques, with their various field-like and Slonczewski-like components \cite{garello2013}, does not enter the analysis. The power of magnetic field-based techniques has been already demonstrated by observing equi-speed contours in Pt/Co/Pt trilayers \cite{Choe}. It has also been suggested that the detection of the Walker breakdown can be used as a direct measure of the DMI \cite{Thiaville_DMI}. However, it is experimentally very difficult to observe the Walker breakdown field due to the fact that it is often not reached or hidden in the creep regime \cite{metaxas}. As will be seen below, the creep regime itself can be used to determine the strength of the DMI.

\begin{figure}[tb]
  \includegraphics[width=8.7cm]{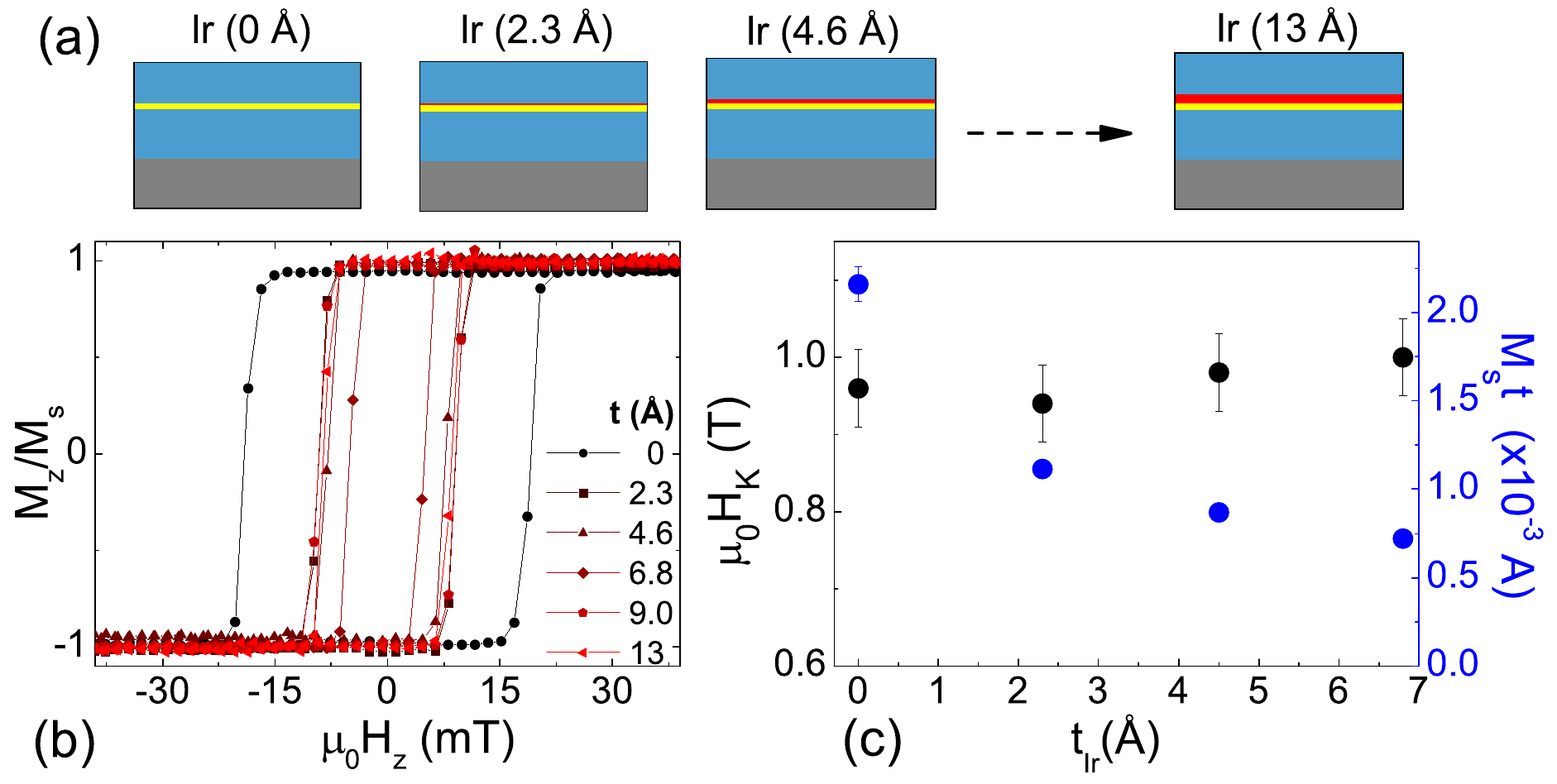}
  \caption{(a) Sketch of the studied Ta/\textcolor{blue2}{Pt}/\textcolor{orange}{Co}/\textcolor{red}{Ir}($t_{\mathrm{Ir}}$)/\textcolor{blue2}{Pt} layer stacks with a varying Ir thickness. (b) Polar Kerr hysteresis loops for samples with various Ir thickness. (c) Anisotropy field $\mu_0 H_\mathrm{K}$ and areal magnetization $M_\mathrm{s}t$ as a function of $t_{\mathrm{Ir}}$. \label{Fig-KerrMicroscopy}}
\end{figure}

The multilayers for our study were grown by room temperature dc sputtering at base pressures $\lesssim10^{-7}$~mbar on thermally oxidized Si substrates with a 3~nm thick Ta buffer layer. In order to reveal the effect of an Ir interface, we started from a stack of Pt(5~nm)/Co(0.7~nm)/Pt(3~nm) and inserted a thin layer of various Ir thicknesses $t_{\mathrm{Ir}}$ at the interface between the Co and top Pt layer, as depicted in Fig.~\ref{Fig-KerrMicroscopy}(a). The films were consequently studied using polar Kerr microscopy. All the films exhibit a perpendicular anisotropy, as shown by the square OOP hysteresis loops presented in Fig.~\ref{Fig-KerrMicroscopy}(b). The coercive field of about 20~mT in Pt/Co/Pt drops to about 9~mT as soon as the top surface is dusted with any thickness of Ir. The OOP anisotropy was measured by the vibrating sample magnetometry technique in an in-plane field configuration. 
Fig.~\ref{Fig-KerrMicroscopy}(c) shows that the anisotropy field $\mu_0K_\mathrm{K}$ is about 1~T for all the films, which demonstrates that the anisotropy comes mostly from the bottom Pt/Co interface \cite{Lacour}. This is experimentally convenient, since it permits us to study changes in the DMI from the inclusion of the Ir layer without the complication of varying OOP anisotropy---and quantities that depend on it such as DW width---also varying.

\begin{figure}[tb]
  \includegraphics[width=7.5cm]{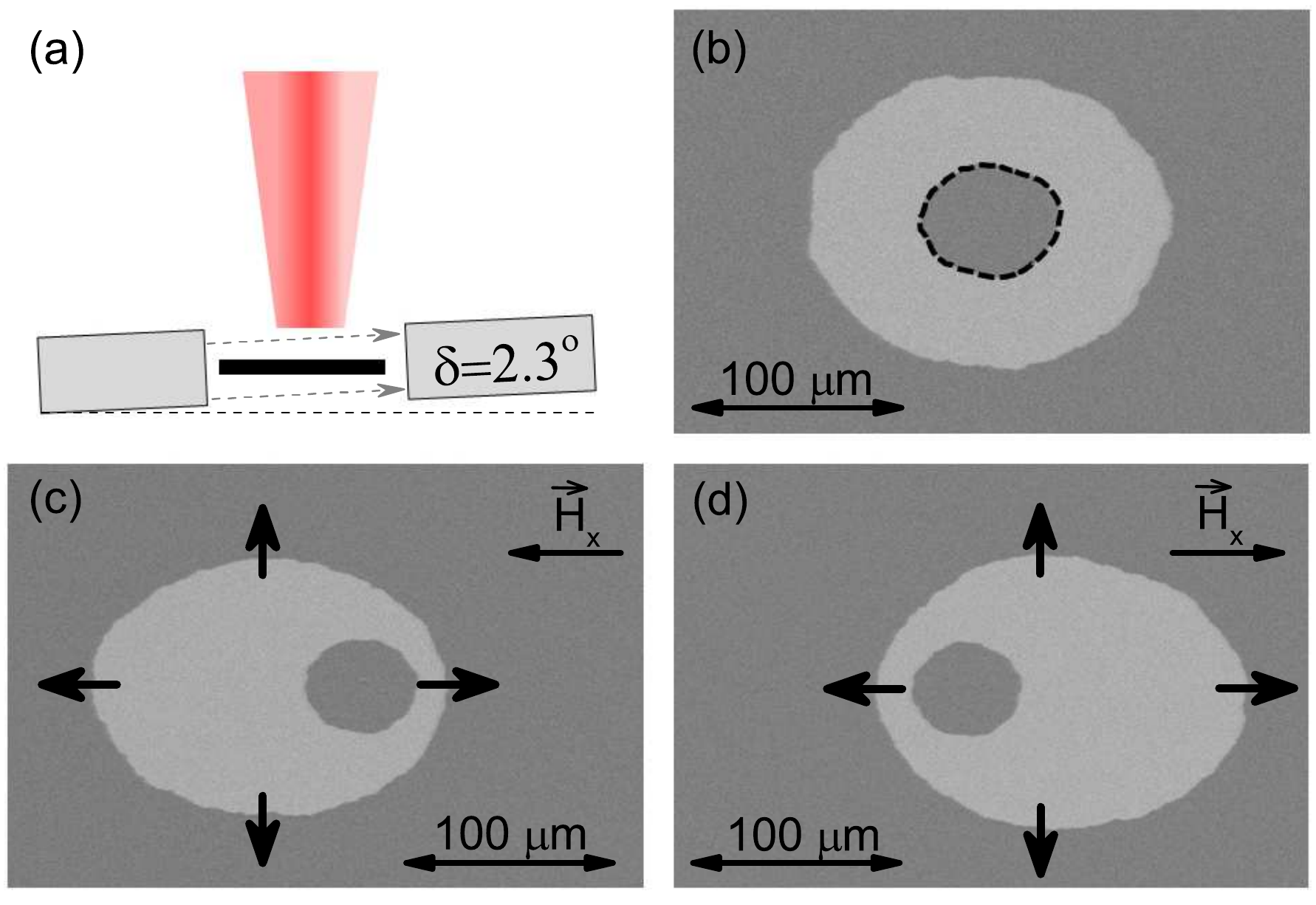}
  \caption{(a) Experimental setup in the Kerr microscope for DMI measurement. The magnetic field is tilted by a small angle $\delta$ with respect to the sample plane. (b) DW displacement in the case of $\delta=90^\circ$ after the application of a 1~s and $\mu_0H_z=7$~mT pulse. The initial DW position is indicated by the dashed line. (c) DW displacement in the case of $\delta=2.3^\circ$ after the application of a 1~s and $\mu_0H_x=+60$~mT pulse. (d) DW displacement after the application of a magnetic field pulse of the same length and $\mu_0H_x=-60$~mT. White arrows indicate the initial orientation of the magnetic moments within the DW. \label{Fig_2}}
\end{figure}

The field-induced DW displacement was investigated by Kerr microscopy in the polar configuration. The experimental setup is shown in Fig.~\ref{Fig_2}(a). The magnetic field was applied in-plane with a small out-of-plane component. This is achieved by tilting the magnet by an angle $\delta$ with respect to the sample plane. This was needed due to the fact that an in-plane field alone is unable to move the DW. The role of the in-plane field is demonstrated in Fig.~\ref{Fig_2}(b)-(d). In each case, a reverse domain was nucleated and allowed to expand a little before switching off the field. Its shape was then recorded, indicated by the dashed line shown in Fig.~\ref{Fig_2}(b). Consequently we applied a 0.8-120~s long pulse of a magnetic field up to 350~mT, during which the domain expands as the DW propagates outwards. In the case of OOP field, i.e. $\delta=90^\circ$, the domain expansion is homogeneous (Fig.~\ref{Fig_2}(b)). The situation was very different in the case of an in-plane field component when $\delta \approx 2.3^\circ$, as shown in Fig.~\ref{Fig_2}(c). One can immediately see that the DW moving to the left and to the right moved with different velocities while the DWs moving in the directions perpendicular to the in-plane field moved with the same velocities. Our explanation for this observation is that the magnetic film contains N\'{e}el walls rather than Bloch walls. The in-plane magnetic field thus breaks the symmetry, and the magnetic moments within the DW on the right would be initially antiparallel, whereas the ones on the left parallel, to the magnetic field. To confirm this hypothesis, we have reversed the the sense of the in-plane magnetic field. The DW displacement after such a magnetic field pulse is shown in Fig.~\ref{Fig_2}(d) with the corresponding initial magnetic moment orientation within the DW.

\begin{figure}[tb]
  \includegraphics[width=7.3cm]{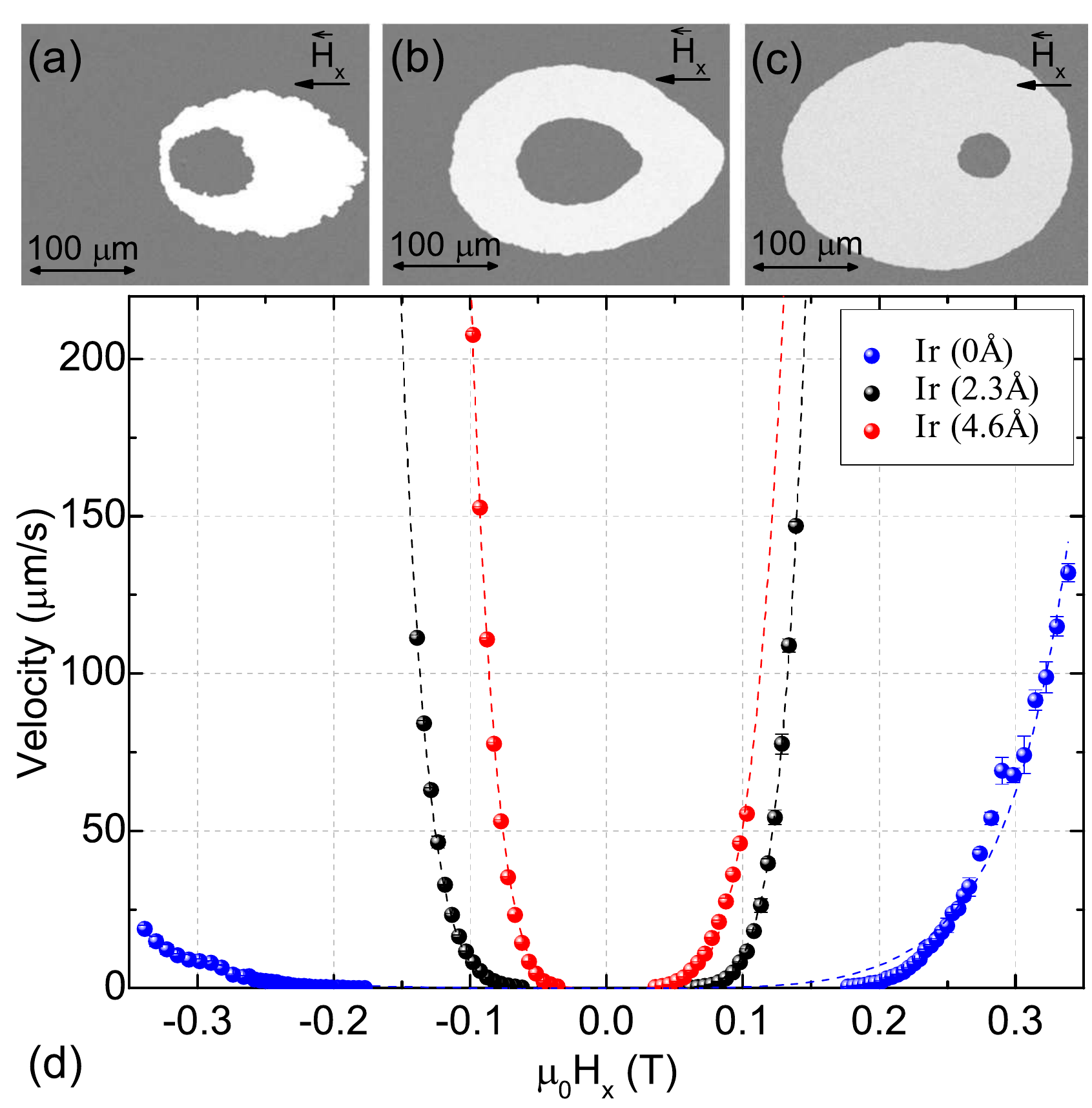}
  \caption{Differential Kerr image of the DW displacement in the case of (a) Ir~(0~\AA), (b) Ir~(2.3~\AA) and (c) Ir~(4.6~\AA) after the application of a 1~s and 320~mT, 1~s and 130~mT, 5~s and 60~mT field pulse, respectively.(d) DW velocity as a function of in-plane magnetic field $H_x$ in the case of Ir~(0~\AA), Ir~(2.3~\AA) and Ir~(4.6~\AA) for a DW creeping along the $x$ direction. The dashed curves show the fits of the creep model described by equation (\ref{equ_creep}) to the data. \label{Fig_3}}
\end{figure}

The average DW velocity during a field pulse can be straightforwardly determined from the DW displacement and the pulse duration. We investigated systematically the DW velocities in the direction of in-plane magnetic field as a function of field pulse strength. A representative picture of the DW motion in a Pt/Co/Pt film is shown in Fig.~\ref{Fig_3}(a), showing the right-hand DW moving much faster than the left-hand one for a left pointing in-plane field component. We emphasise that the DW creep is driven by the small OOP component and the in-plane field component breaks the radial symmetry of the creep velocity. This is expressed by the asymmetry of the velocity-field curves in Fig.~\ref{Fig_3}(d). The detected asymmetry almost disappears in the film with 2.3~\AA\ of Ir (Fig.~\ref{Fig_3}(b)), and has the opposite sign in the samples with no Ir (Fig.~\ref{Fig_3}(a)) and 4.6~\AA\ of Ir (Fig.~\ref{Fig_3}(c)). The corresponding curves in Fig.~\ref{Fig_3}(d) reflect these asymmetries. The inverted asymmetry suggests an inversion of the spin texture within the DWs.


The DW displacement at low magnetic fields follows the creep law \cite{metaxas}, which can be expressed as
\begin{equation}\label{equ_creep}
  v = v_0 \exp\left[-\zeta \left(\mu_0H_z\right)^{-\mu}\right],
\end{equation}
where $\mu=1/4$ is the creep scaling exponent, $v_0$ is the characteristic speed, and $\zeta$ is the scaling coefficient which can be expressed as \cite{Choe}
\begin{equation}
\zeta=\zeta_{0} \left[\sigma(H_x)/\sigma_0 \right]^{1/4},
\end{equation}
where $\zeta_{0}$ is a scaling constant, $\sigma$ is the DW energy density, which is dependent on the in-plane magnetic field $\mu_0H_x$ \cite{Thiaville_DMI}. This dependence can be written as
\begin{equation}
\sigma(H_x)=\sigma_0-\frac{\pi^2\Delta \mu_0^2M_\mathrm{s}^2}{8 K_D} \left(H_x+H_{\mathrm{DMI}}\right)^2
\end{equation}
for the case when the combination of the external magnetic field $\mu_0H_x$ and the intrinsic DM field $\mu_0H_{\mathrm{DMI}}$ is not able to fully transform the Bloch wall into the N\'{e}el wall, i.e. $\mid H_x + H_{\mathrm{DMI}} \mid <4K_D/\pi \mu_0M_\mathrm{s}\equiv \mu_0H_{\mathrm{N-B}}$ and
 \begin{equation}\label{Neel_wall}
\sigma(H_x)=\sigma_0+2K_D\Delta -\pi \Delta \mu_0 M_\mathrm{s} \mid H_x+H_{\mathrm{DMI}}\mid
\end{equation}
in the case of the N\'{e}el wall. In these expressions, $M_\mathrm{s}$ is the saturation magnetization, $\sigma_0$ is the Bloch wall energy density, $K_D$ is the DW anisotropy energy density, and $\Delta$ is the DW width. In this model we use $M_\mathrm{s}=1.1\times10^6$~A/m$^2$, $A=16$~pJ/m, $K_0=\mu_0 (H_\mathrm{K}M_\mathrm{s} - M_\mathrm{s}^2/2) = 3.4\times10^5$~J/m$^3$, $\Delta=\sqrt{A/K_0}=7.2$~nm and $\sigma_0 = 2\pi\sqrt{A K_0}=14$~mJ/m$^2$. The magnetostatic shape anisotropy term favoring the Bloch wall $K_D = N_x\mu_0M_\mathrm{s}^2/2 = 1.7\times10^4$~J/m$^3$ where $N_x$ is the demagnetizing coefficient of the wall \cite{Tarasenko}. As such, this model only requires three fitting parameters that are not determined by other experiments: the scaling parameters $v_0$ and $\zeta_0$, and $H_{\mathrm{DMI}}$ itself. This symmetry-breaking term is thus solely responsible for the asymmetry in the velocity-magnetic field plots.

\begin{figure}[tb]
  \includegraphics[width=8.5cm]{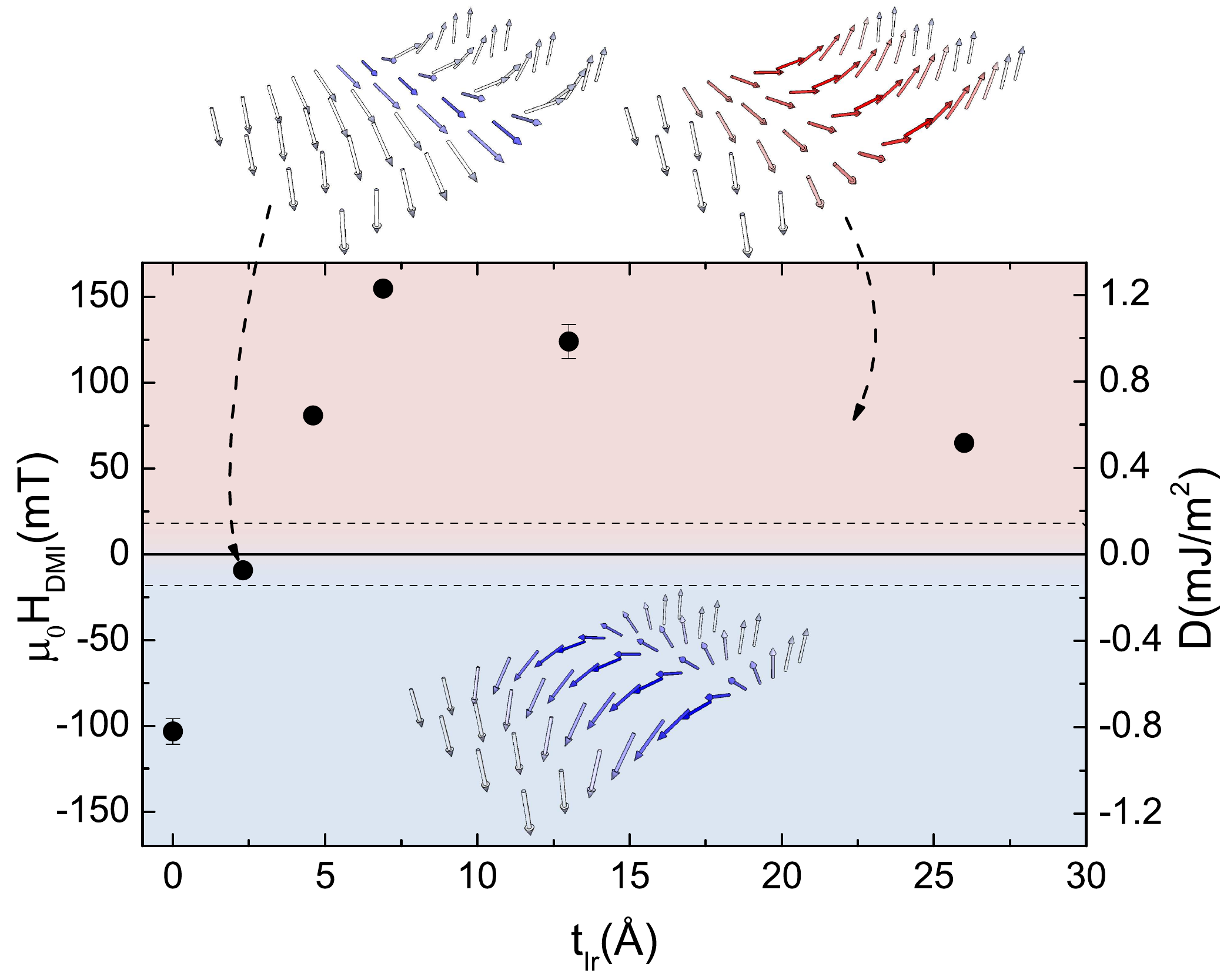}
  \caption{DM field and $D$ as a function of Ir thickness. The region between two dashed lines depicts the range where the DW structure changes continuously from a N\'{e}el wall to a Bloch wall and to a N\'{e}el wall of opposite chirality. Below this line (blue area) right-handed N\'{e}el wall is stable whereas above this line (red area) it is the left-handed N\'{e}el wall. The wall structures are depicted with sketches.}\label{Fig_4}
\end{figure}

This model was fitted to the data for all our samples, with the fitted curves shown as the dashed lines in Fig.~\ref{Fig_3}(d), and the model can be seen to give an excellent description of the experimental results. The extracted DM fields as a function of Ir thickness are displayed in Fig.~\ref{Fig_4}. One can see that the DM field sign reversal qualitatively agrees with the asymmetry reversal shown in Fig.~\ref{Fig_3}(a)-(c). The DM field is large and negative in the Pt/Co/Pt film, nearly compensated in the case of 2.3~\AA\ of Ir and positive for $t_{\mathrm Ir}$ of 4.6~\AA\ or greater. The calculated critical field separating the N\'{e}el wall stability region from the Bloch-N\'{e}el wall transition region is $|\mu_0H_{\mathrm{N-B}}| \approx 18$~mT. When $H_{\mathrm{DMI}}<-H_{\mathrm{N-B}}$, the DMI is able to stabilize the N\'{e}el wall structure of right-handed chirality, whilst for $H_{\mathrm{DMI}}>+H_{\mathrm{N-B}}$  the stable structure is the left-handed N\'{e}el wall, as depicted in Fig.~\ref{Fig_4}. The region between two dashed lines denotes the transition region in which the DW is continuously distorted from the pure Bloch wall towards the N\'{e}el walls of the appropriate chirality. This behaviour is similar to the one observed in epitaxially grown films by Chen \textit{et al}. \cite{chen2013tailoring}, where the DM constant reverses sign on a similar length scale upon insertion of a thin Ir interlayer. We also emphasize that the suggested DW structure depicted in Fig.~\ref{Fig_2} is no longer valid during the magnetic field pulse and all the magnetic moments eventually reorient into the field direction for sufficiently high magnetic fields. Such DWs, despite the similar magnetic moment orientation, have different energy expressed by equation (\ref{Neel_wall}). This is reflected in different resulting velocities in the creep regime.

We also estimate the effective DM constant $D$ by using the expression $D=\mu_0 H_{\mathrm{DMI}} M_\mathrm{s} \Delta$ \cite{Thiaville_DMI}. This is given on the right-hand ordinate axis of Fig.~\ref{Fig_4}. It is apparent that the DMI in these samples is controlled largely by the top interface, in contrast to the OOP anisotropy, which we saw above to be dominated by the bottom interface. The strongest DMI, $D = 1.2\pm0.1$~mJ, is obtained in the case of Pt/Co/Ir which can be compared to the critical DMI $D_{\mathrm{crit}}$ resulting in a non-uniform magnetization state such as a cycloidal or skyrmionic phase. The critical DM constant can be estimated by using $D_{\mathrm{crit}}=4/\pi\sqrt{AK_0}$ \cite{heide2008dzyaloshinskii}, which in this case is $D_{\mathrm{crit}}\sim 3$~mJ/m$^2$. However, the case of $D<D_{\mathrm{crit}}$ is very important for applications due to the coexistence of ferromagnetic and skyrmionic phases, so that isolated skyrmions can be used for information encoding \cite{sampaio2013nucleation}.

\begin{figure}[tb]
  \includegraphics[width=8.5cm]{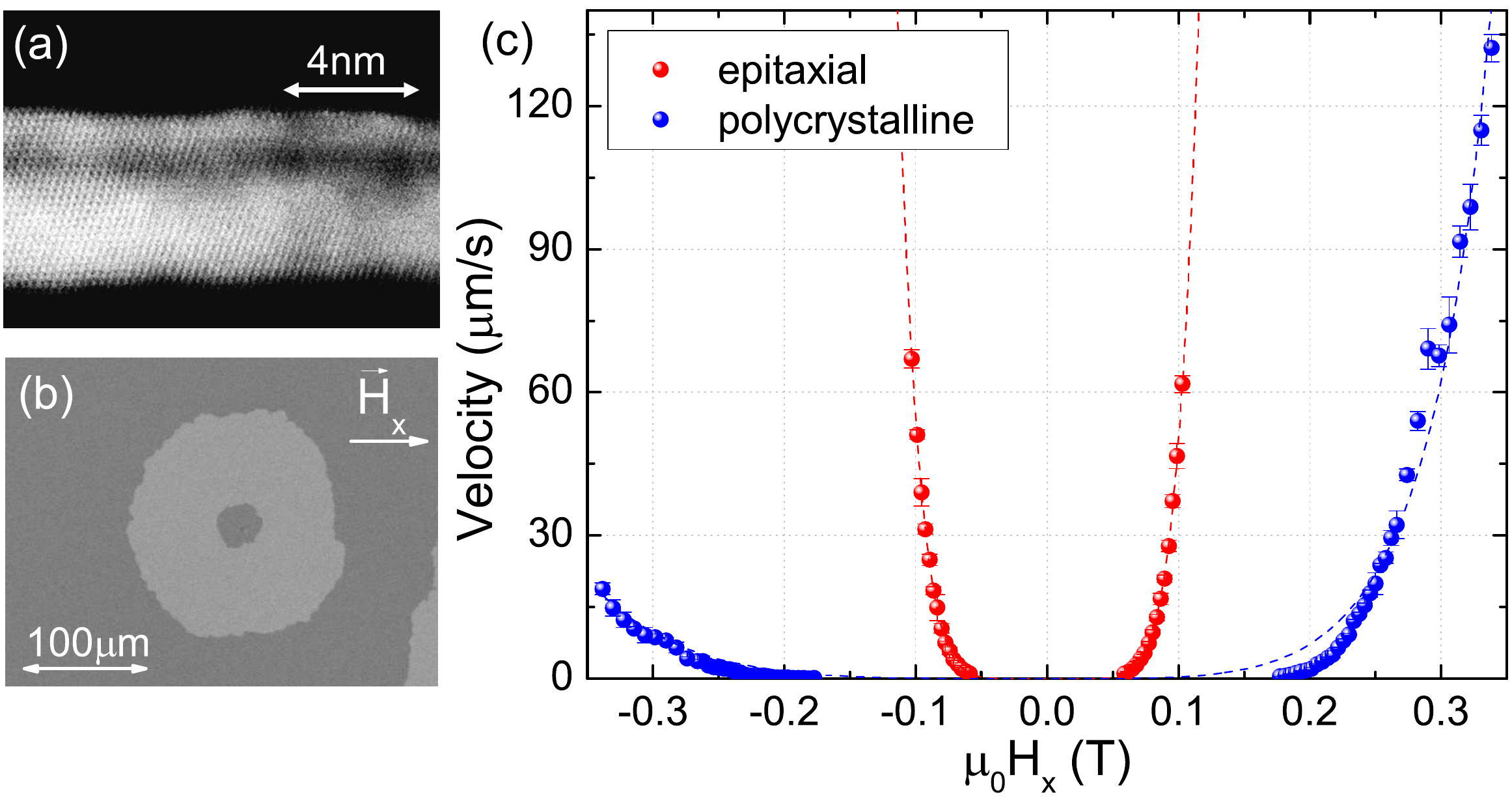}
  \caption{(a) High-angle annular dark-field in scanning transmission electron micrograph of the epitaxial Pt(3~nm)/Co(0.7~nm)/Pt(1~nm) trilayer. The darker Co layer is sandwiched between the two brighter Pt layers. (b) Differential Kerr image of the DW displacement in the epitaxial Co/Pt/Co sample after the application of a 1~s long, $\mu_0H_x=100$~mT field pulse. (c) Comparison of DW velocities as a function of magnetic field in the polycrystalline and epitaxial films. The dashed curves show the fits of the creep model described by equation (\ref{equ_creep}).}\label{Fig_5}
\end{figure}

A strong DMI is also measured in the most structurally symmetric sample of Pt/Co/Pt, where one would not expect any DMI at all. In order to understand the origin of the strong DMI in the stack of Pt/Co/Pt, we grew a similar stack of Pt(3~nm)/Co(0.7~nm)/Pt(1~nm) epitaxially. The seed Pt layer was grown by the sputtering technique on a C-plane sapphire substrate at $500^\circ$C followed by the Co/Pt bilayer sputtering at $100^\circ$C, as described in Ref. \onlinecite{Mihai}. The epitaxial character of the grown film was confirmed by X-ray diffraction and high-angle annular dark-field imaging in a scanning transmission electron microscope. Fig.~\ref{Fig_5}(a) shows the high level of crystallographic ordering in the epitaxial trilayer. In order to study the DMI we have performed the same measurements as described above and Fig.~\ref{Fig_5}(b) shows a representative DW displacement for the epitaxial sample. One can directly see the striking difference from the picture obtained on the polycrystalline Ta/Pt/Co/Pt sample that was shown in Fig.~\ref{Fig_3}(a). The observed asymmetry is in this case suppressed and the DW displacement becomes radially symmetric. This is also expressed by the symmetric velocity-field curve shown in Fig.~\ref{Fig_5}(c) resulting in $D = 0.02\pm0.01$~mJ. The effective DMI thus vanishes in the case of the crystallographically symmetric interfaces on either side of the ferromagnet, just as expected. An important conclusion from the demonstrated experiment is that the DMI shows exquisite sensitivity to the atomic-scale details of the interfacial structure in these kinds of multilayer. Nevertheless, characterising the details of potentially asymmetric interface properties such as the roughness, degree of intermixing, density of stacking faults, remains an outstanding materials science challenge.

Besides the asymmetric metal composition and crystallographic structure around the ferromagnetic layer, the asymmetrically induced magnetic moment may play an important role. It has been shown that Pt and Ir exhibit strong proximity effect in the vicinity of a ferromagnet \cite{Fisher_XMCD} therefore one would expect different induced magnetic moment on either side of the Co layer. In our magnetometry data shown in Fig.~\ref{Fig-KerrMicroscopy}(c) we see a significant drop of normalized magnetization once the Ir layer is inserted between the top Co/Pt interface indicating a decrease of induced magnetic moment in the top layer. The effect of this asymmetry on the DMI is not yet known.

In conclusion, we have demonstrated a simple-to-implement magnetic field-based method for the DMI detection and measurement in out-of-plane anisotropy materials. The DMI was quantified \textit{ex situ} by Kerr microscopy in sputtered Pt/Co/Ir/Pt layers. We are able to control the DW chirality by changing the thickness of Ir film via in inversion of the effective intrinsic DM field. We also reveal the crucial importance of the exact nature of the ferromagnet/heavy metal interface for the DMI by comparing a polycrystalline multilayer of the type studied in most laboratories to a similar multilayer with controlled crystallographic order. The method we present opens the way for fast and convenient exploration of the DMI in new multilayer structures intended for use in DW and skyrmion racetrack memories.

\begin{acknowledgments}
This work was supported by the UK EPSRC (grant numbers EP/I011668/1, EP/I013520/1, EP/K003127/1 and EP/J007110/1), the Scottish Universities Physics Alliance and the University of Glasgow. The authors thank to Stefania Pizzini for helpful discussion.
\end{acknowledgments}


\begin{thebibliography}{42}
\expandafter\ifx\csname natexlab\endcsname\relax\def\natexlab#1{#1}\fi
\expandafter\ifx\csname bibnamefont\endcsname\relax
  \def\bibnamefont#1{#1}\fi
\expandafter\ifx\csname bibfnamefont\endcsname\relax
  \def\bibfnamefont#1{#1}\fi
\expandafter\ifx\csname citenamefont\endcsname\relax
  \def\citenamefont#1{#1}\fi
\expandafter\ifx\csname url\endcsname\relax
  \def\url#1{\texttt{#1}}\fi
\expandafter\ifx\csname urlprefix\endcsname\relax\def\urlprefix{URL }\fi
\providecommand{\bibinfo}[2]{#2}
\providecommand{\eprint}[2][]{\url{#2}}

\bibitem[{\citenamefont{Dzyaloshinsky}(1958)}]{dzyaloshinsky1958}
\bibinfo{author}{\bibfnamefont{I.}~\bibnamefont{Dzyaloshinsky}},
  \bibinfo{journal}{J. Phys. Chem. Solids} \textbf{\bibinfo{volume}{4}},
  \bibinfo{pages}{241} (\bibinfo{year}{1958}).

\bibitem[{\citenamefont{Moriya}(1960)}]{moriya1960}
\bibinfo{author}{\bibfnamefont{T.}~\bibnamefont{Moriya}},
  \bibinfo{journal}{Phys. Rev.} \textbf{\bibinfo{volume}{120}},
  \bibinfo{pages}{91} (\bibinfo{year}{1960}).

\bibitem[{\citenamefont{Gan et~al.}(2000)\citenamefont{Gan, Chung, Aschenbach,
  Dreyer, and Gomez}}]{ganieee2000}
\bibinfo{author}{\bibfnamefont{L.}~\bibnamefont{Gan}},
  \bibinfo{author}{\bibfnamefont{S.~H.} \bibnamefont{Chung}},
  \bibinfo{author}{\bibfnamefont{K.~H.} \bibnamefont{Aschenbach}},
  \bibinfo{author}{\bibfnamefont{M.}~\bibnamefont{Dreyer}}, \bibnamefont{and}
  \bibinfo{author}{\bibfnamefont{R.~D.} \bibnamefont{Gomez}},
  \bibinfo{journal}{IEEE Trans. Magn.} \textbf{\bibinfo{volume}{36}},
  \bibinfo{pages}{3047} (\bibinfo{year}{2000}).

\bibitem[{\citenamefont{Kl{\"a}ui et~al.}(2005)\citenamefont{Kl{\"a}ui, Vaz,
  Bland, Wernsdorfer, Faini, Cambril, Heyderman, Nolting, and
  R{\"u}diger}}]{klauiprl2005}
\bibinfo{author}{\bibfnamefont{M.}~\bibnamefont{Kl{\"a}ui}},
  \bibinfo{author}{\bibfnamefont{C.~A.~F.} \bibnamefont{Vaz}},
  \bibinfo{author}{\bibfnamefont{J.~A.~C.} \bibnamefont{Bland}},
  \bibinfo{author}{\bibfnamefont{W.}~\bibnamefont{Wernsdorfer}},
  \bibinfo{author}{\bibfnamefont{G.}~\bibnamefont{Faini}},
  \bibinfo{author}{\bibfnamefont{E.}~\bibnamefont{Cambril}},
  \bibinfo{author}{\bibfnamefont{L.~J.} \bibnamefont{Heyderman}},
  \bibinfo{author}{\bibfnamefont{F.}~\bibnamefont{Nolting}}, \bibnamefont{and}
  \bibinfo{author}{\bibfnamefont{U.}~\bibnamefont{R{\"u}diger}},
  \bibinfo{journal}{Phys. Rev. Lett.} \textbf{\bibinfo{volume}{94}},
  \bibinfo{pages}{106601} (\bibinfo{year}{2005}).

\bibitem[{\citenamefont{Yamaguchi et~al.}(2004)\citenamefont{Yamaguchi, Ono,
  Nasu, Miyake, Mibu, and Shinjo}}]{yamaguchiprl2004}
\bibinfo{author}{\bibfnamefont{A.}~\bibnamefont{Yamaguchi}},
  \bibinfo{author}{\bibfnamefont{T.}~\bibnamefont{Ono}},
  \bibinfo{author}{\bibfnamefont{S.}~\bibnamefont{Nasu}},
  \bibinfo{author}{\bibfnamefont{K.}~\bibnamefont{Miyake}},
  \bibinfo{author}{\bibfnamefont{K.}~\bibnamefont{Mibu}}, \bibnamefont{and}
  \bibinfo{author}{\bibfnamefont{T.}~\bibnamefont{Shinjo}},
  \bibinfo{journal}{Phys. Rev. Lett.} \textbf{\bibinfo{volume}{92}},
  \bibinfo{pages}{077205} (\bibinfo{year}{2004}).

\bibitem[{\citenamefont{Hayashi et~al.}(2006)\citenamefont{Hayashi, Thomas,
  Rettner, Moriya, and Parkin}}]{hayashi2007}
\bibinfo{author}{\bibfnamefont{M.}~\bibnamefont{Hayashi}},
  \bibinfo{author}{\bibfnamefont{L.}~\bibnamefont{Thomas}},
  \bibinfo{author}{\bibfnamefont{C.}~\bibnamefont{Rettner}},
  \bibinfo{author}{\bibfnamefont{R.}~\bibnamefont{Moriya}}, \bibnamefont{and}
  \bibinfo{author}{\bibfnamefont{S.~S.~P.} \bibnamefont{Parkin}},
  \bibinfo{journal}{Nature Phys.} \textbf{\bibinfo{volume}{3}},
  \bibinfo{pages}{21} (\bibinfo{year}{2006}).

\bibitem[{\citenamefont{Meier et~al.}(2007)\citenamefont{Meier, Bolte, Eiselt,
  Kr\"{u}ger, Kim, and Fischer}}]{meier2007}
\bibinfo{author}{\bibfnamefont{G.}~\bibnamefont{Meier}},
  \bibinfo{author}{\bibfnamefont{M.}~\bibnamefont{Bolte}},
  \bibinfo{author}{\bibfnamefont{R.}~\bibnamefont{Eiselt}},
  \bibinfo{author}{\bibfnamefont{B.}~\bibnamefont{Kr\"{u}ger}},
  \bibinfo{author}{\bibfnamefont{D.~H.} \bibnamefont{Kim}}, \bibnamefont{and}
  \bibinfo{author}{\bibfnamefont{P.}~\bibnamefont{Fischer}},
  \bibinfo{journal}{Phys. Rev. Lett.} \textbf{\bibinfo{volume}{98}},
  \bibinfo{pages}{187202} (\bibinfo{year}{2007}).

\bibitem[{\citenamefont{Lepadatu et~al.}(2009)\citenamefont{Lepadatu,
  Vanhaverbeke, Atkinson, Allenspach, and Marrows}}]{lepadatu2009}
\bibinfo{author}{\bibfnamefont{S.}~\bibnamefont{Lepadatu}},
  \bibinfo{author}{\bibfnamefont{A.}~\bibnamefont{Vanhaverbeke}},
  \bibinfo{author}{\bibfnamefont{D.}~\bibnamefont{Atkinson}},
  \bibinfo{author}{\bibfnamefont{R.}~\bibnamefont{Allenspach}},
  \bibnamefont{and} \bibinfo{author}{\bibfnamefont{C.~H.}
  \bibnamefont{Marrows}}, \bibinfo{journal}{Phys. Rev. Lett.}
  \textbf{\bibinfo{volume}{102}}, \bibinfo{pages}{127203}
  (\bibinfo{year}{2009}).

\bibitem[{\citenamefont{Boulle et~al.}(2008)\citenamefont{Boulle, Kimling,
  Warnicke, Kl\"{a}ui, R\"{u}diger, Malinowski, Swagten, Koopmans, Ulysse, and
  Faini}}]{boulle2008}
\bibinfo{author}{\bibfnamefont{O.}~\bibnamefont{Boulle}},
  \bibinfo{author}{\bibfnamefont{J.}~\bibnamefont{Kimling}},
  \bibinfo{author}{\bibfnamefont{P.}~\bibnamefont{Warnicke}},
  \bibinfo{author}{\bibfnamefont{M.}~\bibnamefont{Kl\"{a}ui}},
  \bibinfo{author}{\bibfnamefont{U.}~\bibnamefont{R\"{u}diger}},
  \bibinfo{author}{\bibfnamefont{G.}~\bibnamefont{Malinowski}},
  \bibinfo{author}{\bibfnamefont{H.~J.} \bibnamefont{Swagten}},
  \bibinfo{author}{\bibfnamefont{B.}~\bibnamefont{Koopmans}},
  \bibinfo{author}{\bibfnamefont{C.}~\bibnamefont{Ulysse}}, \bibnamefont{and}
  \bibinfo{author}{\bibfnamefont{G.}~\bibnamefont{Faini}},
  \bibinfo{journal}{Phys. Rev. Lett.} \textbf{\bibinfo{volume}{101}},
  \bibinfo{pages}{216601} (\bibinfo{year}{2008}).

\bibitem[{\citenamefont{{San Emeterio Alvarez} et~al.}(2010)\citenamefont{{San
  Emeterio Alvarez}, Wang, Lepadatu, Landi, Bending, and
  Marrows}}]{alvarezprl2010}
\bibinfo{author}{\bibfnamefont{L.}~\bibnamefont{{San Emeterio Alvarez}}},
  \bibinfo{author}{\bibfnamefont{K.-Y.} \bibnamefont{Wang}},
  \bibinfo{author}{\bibfnamefont{S.}~\bibnamefont{Lepadatu}},
  \bibinfo{author}{\bibfnamefont{S.}~\bibnamefont{Landi}},
  \bibinfo{author}{\bibfnamefont{S.~J.} \bibnamefont{Bending}},
  \bibnamefont{and} \bibinfo{author}{\bibfnamefont{C.~H.}
  \bibnamefont{Marrows}}, \bibinfo{journal}{Phys. Rev. Lett.}
  \textbf{\bibinfo{volume}{104}}, \bibinfo{pages}{137205}
  (\bibinfo{year}{2010}).

\bibitem[{\citenamefont{Moore et~al.}(2008)\citenamefont{Moore, Miron, Gaudin,
  Serret, Auffret, Rodmacq, Schuhl, Pizzini, Vogel, and
  Bonfim}}]{Moore_APL2008}
\bibinfo{author}{\bibfnamefont{T.~A.} \bibnamefont{Moore}},
  \bibinfo{author}{\bibfnamefont{I.~M.} \bibnamefont{Miron}},
  \bibinfo{author}{\bibfnamefont{G.}~\bibnamefont{Gaudin}},
  \bibinfo{author}{\bibfnamefont{G.}~\bibnamefont{Serret}},
  \bibinfo{author}{\bibfnamefont{S.}~\bibnamefont{Auffret}},
  \bibinfo{author}{\bibfnamefont{B.}~\bibnamefont{Rodmacq}},
  \bibinfo{author}{\bibfnamefont{A.}~\bibnamefont{Schuhl}},
  \bibinfo{author}{\bibfnamefont{S.}~\bibnamefont{Pizzini}},
  \bibinfo{author}{\bibfnamefont{J.}~\bibnamefont{Vogel}}, \bibnamefont{and}
  \bibinfo{author}{\bibfnamefont{M.}~\bibnamefont{Bonfim}},
  \bibinfo{journal}{Appl. Phys. Lett.} \textbf{\bibinfo{volume}{93}},
  \bibinfo{pages}{262504} (\bibinfo{year}{2008}).

\bibitem[{\citenamefont{Miron et~al.}(2010)\citenamefont{Miron, Gaudin,
  Auffret, Rodmacq, Schuhl, Pizzini, Vogel, and
  Gambardella}}]{Miron_Nature2010}
\bibinfo{author}{\bibfnamefont{I.~M.} \bibnamefont{Miron}},
  \bibinfo{author}{\bibfnamefont{G.}~\bibnamefont{Gaudin}},
  \bibinfo{author}{\bibfnamefont{S.}~\bibnamefont{Auffret}},
  \bibinfo{author}{\bibfnamefont{B.}~\bibnamefont{Rodmacq}},
  \bibinfo{author}{\bibfnamefont{A.}~\bibnamefont{Schuhl}},
  \bibinfo{author}{\bibfnamefont{S.}~\bibnamefont{Pizzini}},
  \bibinfo{author}{\bibfnamefont{J.}~\bibnamefont{Vogel}}, \bibnamefont{and}
  \bibinfo{author}{\bibfnamefont{P.}~\bibnamefont{Gambardella}},
  \bibinfo{journal}{Nature Mat.} \textbf{\bibinfo{volume}{9}},
  \bibinfo{pages}{230} (\bibinfo{year}{2010}).

\bibitem[{\citenamefont{Liu et~al.}(2012)\citenamefont{Liu, Pai, Li, Tseng,
  Ralph, and Buhrman}}]{Buhrman_Science}
\bibinfo{author}{\bibfnamefont{L.}~\bibnamefont{Liu}},
  \bibinfo{author}{\bibfnamefont{C.-F.} \bibnamefont{Pai}},
  \bibinfo{author}{\bibfnamefont{Y.}~\bibnamefont{Li}},
  \bibinfo{author}{\bibfnamefont{H.~W.} \bibnamefont{Tseng}},
  \bibinfo{author}{\bibfnamefont{D.~C.} \bibnamefont{Ralph}}, \bibnamefont{and}
  \bibinfo{author}{\bibfnamefont{R.~A.} \bibnamefont{Buhrman}},
  \bibinfo{journal}{Science} \textbf{\bibinfo{volume}{336}},
  \bibinfo{pages}{555} (\bibinfo{year}{2012}).

\bibitem[{\citenamefont{Malozemoff and
  Slonczewski}(1979)}]{malozemoff1979magnetic}
\bibinfo{author}{\bibfnamefont{A.}~\bibnamefont{Malozemoff}} \bibnamefont{and}
  \bibinfo{author}{\bibfnamefont{J.}~\bibnamefont{Slonczewski}},
  \emph{\bibinfo{title}{Magnetic domain walls in bubble materials}},
  vol.~\bibinfo{volume}{81} (\bibinfo{publisher}{Academic Press New York},
  \bibinfo{year}{1979}).

\bibitem[{\citenamefont{Khvalkovskiy et~al.}(2013)\citenamefont{Khvalkovskiy,
  Cros, Apalkov, Nikitin, Krounbi, Zvezdin, Anane, Grollier, and
  Fert}}]{khvalkovskiy2013matching}
\bibinfo{author}{\bibfnamefont{A.~V.} \bibnamefont{Khvalkovskiy}},
  \bibinfo{author}{\bibfnamefont{V.}~\bibnamefont{Cros}},
  \bibinfo{author}{\bibfnamefont{D.}~\bibnamefont{Apalkov}},
  \bibinfo{author}{\bibfnamefont{V.}~\bibnamefont{Nikitin}},
  \bibinfo{author}{\bibfnamefont{M.}~\bibnamefont{Krounbi}},
  \bibinfo{author}{\bibfnamefont{K.}~\bibnamefont{Zvezdin}},
  \bibinfo{author}{\bibfnamefont{A.}~\bibnamefont{Anane}},
  \bibinfo{author}{\bibfnamefont{J.}~\bibnamefont{Grollier}}, \bibnamefont{and}
  \bibinfo{author}{\bibfnamefont{A.}~\bibnamefont{Fert}},
  \bibinfo{journal}{Phys. Rev. B} \textbf{\bibinfo{volume}{87}},
  \bibinfo{pages}{020402} (\bibinfo{year}{2013}).

\bibitem[{\citenamefont{Haazen et~al.}(2013)\citenamefont{Haazen, Mur{\`e},
  Franken, Lavrijsen, Swagten, and Koopmans}}]{haazen2013domain}
\bibinfo{author}{\bibfnamefont{P.~P.~J.} \bibnamefont{Haazen}},
  \bibinfo{author}{\bibfnamefont{E.}~\bibnamefont{Mur{\`e}}},
  \bibinfo{author}{\bibfnamefont{J.~H.} \bibnamefont{Franken}},
  \bibinfo{author}{\bibfnamefont{R.}~\bibnamefont{Lavrijsen}},
  \bibinfo{author}{\bibfnamefont{H.~J.~M.} \bibnamefont{Swagten}},
  \bibnamefont{and} \bibinfo{author}{\bibfnamefont{B.}~\bibnamefont{Koopmans}},
  \bibinfo{journal}{Nature Mat.} \textbf{\bibinfo{volume}{12}},
  \bibinfo{pages}{299} (\bibinfo{year}{2013}).

\bibitem[{\citenamefont{Martinez et~al.}(2013)\citenamefont{Martinez, Emori,
  and Beach}}]{Martinez_APL}
\bibinfo{author}{\bibfnamefont{E.}~\bibnamefont{Martinez}},
  \bibinfo{author}{\bibfnamefont{S.}~\bibnamefont{Emori}}, \bibnamefont{and}
  \bibinfo{author}{\bibfnamefont{G.~S.~D.} \bibnamefont{Beach}},
  \bibinfo{journal}{Appl. Phys. Lett.} \textbf{\bibinfo{volume}{103}},
  \bibinfo{pages}{072406} (\bibinfo{year}{2013}).

\bibitem[{\citenamefont{Parkin et~al.}(2008)\citenamefont{Parkin, Hayashi, and
  Thomas}}]{parkin2008}
\bibinfo{author}{\bibfnamefont{S.~S.~P.} \bibnamefont{Parkin}},
  \bibinfo{author}{\bibfnamefont{M.}~\bibnamefont{Hayashi}}, \bibnamefont{and}
  \bibinfo{author}{\bibfnamefont{L.}~\bibnamefont{Thomas}},
  \bibinfo{journal}{Science} \textbf{\bibinfo{volume}{320}},
  \bibinfo{pages}{190} (\bibinfo{year}{2008}).

\bibitem[{\citenamefont{Thiaville et~al.}(2012)\citenamefont{Thiaville, Rohart,
  Ju\'{e}, Cros, and Fert}}]{Thiaville_DMI}
\bibinfo{author}{\bibfnamefont{A.}~\bibnamefont{Thiaville}},
  \bibinfo{author}{\bibfnamefont{S.}~\bibnamefont{Rohart}},
  \bibinfo{author}{\bibfnamefont{E.}~\bibnamefont{Ju\'{e}}},
  \bibinfo{author}{\bibfnamefont{V.}~\bibnamefont{Cros}}, \bibnamefont{and}
  \bibinfo{author}{\bibfnamefont{A.}~\bibnamefont{Fert}},
  \bibinfo{journal}{Europhys. Lett.} \textbf{\bibinfo{volume}{100}},
  \bibinfo{eid}{57002} (\bibinfo{year}{2012}).

\bibitem[{\citenamefont{Kim et~al.}(2013)\citenamefont{Kim, Lee, Lee, and
  Stiles}}]{Kim_ChiralityFromSOI}
\bibinfo{author}{\bibfnamefont{K.-W.} \bibnamefont{Kim}},
  \bibinfo{author}{\bibfnamefont{H.-W.} \bibnamefont{Lee}},
  \bibinfo{author}{\bibfnamefont{K.-J.} \bibnamefont{Lee}}, \bibnamefont{and}
  \bibinfo{author}{\bibfnamefont{M.~D.} \bibnamefont{Stiles}},
  \bibinfo{journal}{Phys. Rev. Lett.} \textbf{\bibinfo{volume}{111}},
  \bibinfo{pages}{216601} (\bibinfo{year}{2013}).

\bibitem[{\citenamefont{Emori et~al.}(2013)\citenamefont{Emori, Bauer, Ahn,
  Martinez, and Beach}}]{emori2013current}
\bibinfo{author}{\bibfnamefont{S.}~\bibnamefont{Emori}},
  \bibinfo{author}{\bibfnamefont{U.}~\bibnamefont{Bauer}},
  \bibinfo{author}{\bibfnamefont{S.-M.} \bibnamefont{Ahn}},
  \bibinfo{author}{\bibfnamefont{E.}~\bibnamefont{Martinez}}, \bibnamefont{and}
  \bibinfo{author}{\bibfnamefont{G.~S.~D.} \bibnamefont{Beach}},
  \bibinfo{journal}{Nature Mat.} \textbf{\bibinfo{volume}{12}},
  \bibinfo{pages}{611} (\bibinfo{year}{2013}).

\bibitem[{\citenamefont{Ryu et~al.}(2013)\citenamefont{Ryu, Thomas, Yang, and
  Parkin}}]{ryu2013chiral}
\bibinfo{author}{\bibfnamefont{K.-S.} \bibnamefont{Ryu}},
  \bibinfo{author}{\bibfnamefont{L.}~\bibnamefont{Thomas}},
  \bibinfo{author}{\bibfnamefont{S.-H.} \bibnamefont{Yang}}, \bibnamefont{and}
  \bibinfo{author}{\bibfnamefont{S.}~\bibnamefont{Parkin}},
  \bibinfo{journal}{Nature Nano.} \textbf{\bibinfo{volume}{8}},
  \bibinfo{pages}{527} (\bibinfo{year}{2013}).

\bibitem[{\citenamefont{Lee et~al.}(2014)\citenamefont{Lee, Liu, Pai, Li,
  Tseng, Gowtham, Park, Ralph, and Buhrman}}]{lee2014}
\bibinfo{author}{\bibfnamefont{O.~J.} \bibnamefont{Lee}},
  \bibinfo{author}{\bibfnamefont{L.~Q.} \bibnamefont{Liu}},
  \bibinfo{author}{\bibfnamefont{C.~F.} \bibnamefont{Pai}},
  \bibinfo{author}{\bibfnamefont{Y.}~\bibnamefont{Li}},
  \bibinfo{author}{\bibfnamefont{H.~W.} \bibnamefont{Tseng}},
  \bibinfo{author}{\bibfnamefont{P.~G.} \bibnamefont{Gowtham}},
  \bibinfo{author}{\bibfnamefont{J.~P.} \bibnamefont{Park}},
  \bibinfo{author}{\bibfnamefont{D.~C.} \bibnamefont{Ralph}}, \bibnamefont{and}
  \bibinfo{author}{\bibfnamefont{R.~A.} \bibnamefont{Buhrman}},
  \bibinfo{journal}{Phys. Rev. B} \textbf{\bibinfo{volume}{89}},
  \bibinfo{pages}{024418} (\bibinfo{year}{2014}).

\bibitem[{\citenamefont{Torrejon et~al.}(2014)\citenamefont{Torrejon, Kim,
  Sinha, Mitani, Hayashi, Yamanouchia, and Ohno}}]{torrejon2014}
\bibinfo{author}{\bibfnamefont{J.}~\bibnamefont{Torrejon}},
  \bibinfo{author}{\bibfnamefont{J.}~\bibnamefont{Kim}},
  \bibinfo{author}{\bibfnamefont{J.}~\bibnamefont{Sinha}},
  \bibinfo{author}{\bibfnamefont{S.}~\bibnamefont{Mitani}},
  \bibinfo{author}{\bibfnamefont{M.}~\bibnamefont{Hayashi}},
  \bibinfo{author}{\bibfnamefont{M.}~\bibnamefont{Yamanouchia}},
  \bibnamefont{and} \bibinfo{author}{\bibfnamefont{H.}~\bibnamefont{Ohno}}
  (\bibinfo{year}{2014}), \eprint{arXiv:1401.3568 [cond-mat.mes-hall]}.

\bibitem[{\citenamefont{Fert}(1991)}]{fert1991magnetic}
\bibinfo{author}{\bibfnamefont{A.}~\bibnamefont{Fert}},
  \bibinfo{journal}{Mater. Sci. Forum} \textbf{\bibinfo{volume}{59}},
  \bibinfo{pages}{439} (\bibinfo{year}{1991}).

\bibitem[{\citenamefont{Cr\'{e}pieux and Lacroix}(1998)}]{crepieux1998}
\bibinfo{author}{\bibfnamefont{A.}~\bibnamefont{Cr\'{e}pieux}}
  \bibnamefont{and} \bibinfo{author}{\bibfnamefont{C.}~\bibnamefont{Lacroix}},
  \bibinfo{journal}{J. Magn. Magn. Mater.} \textbf{\bibinfo{volume}{182}},
  \bibinfo{pages}{341} (\bibinfo{year}{1998}).

\bibitem[{\citenamefont{Pietzsch et~al.}(2004)\citenamefont{Pietzsch, Kubetzka,
  Bode, and Wiesendanger}}]{Kubetzka_2003}
\bibinfo{author}{\bibfnamefont{O.}~\bibnamefont{Pietzsch}},
  \bibinfo{author}{\bibfnamefont{A.}~\bibnamefont{Kubetzka}},
  \bibinfo{author}{\bibfnamefont{M.}~\bibnamefont{Bode}}, \bibnamefont{and}
  \bibinfo{author}{\bibfnamefont{R.}~\bibnamefont{Wiesendanger}},
  \bibinfo{journal}{Phys. Rev. Lett.} \textbf{\bibinfo{volume}{92}},
  \bibinfo{pages}{057202} (\bibinfo{year}{2004}).

\bibitem[{\citenamefont{Bode et~al.}(2007)\citenamefont{Bode, Heide, von
  Bergmann, Ferriani, Heinze, Bihlmayer, Kubetzka, Pietzsch, Bl\"{u}gel, and
  Wiesendanger}}]{Bode}
\bibinfo{author}{\bibfnamefont{M.}~\bibnamefont{Bode}},
  \bibinfo{author}{\bibfnamefont{M.}~\bibnamefont{Heide}},
  \bibinfo{author}{\bibfnamefont{K.}~\bibnamefont{von Bergmann}},
  \bibinfo{author}{\bibfnamefont{P.}~\bibnamefont{Ferriani}},
  \bibinfo{author}{\bibfnamefont{S.}~\bibnamefont{Heinze}},
  \bibinfo{author}{\bibfnamefont{G.}~\bibnamefont{Bihlmayer}},
  \bibinfo{author}{\bibfnamefont{A.}~\bibnamefont{Kubetzka}},
  \bibinfo{author}{\bibfnamefont{O.}~\bibnamefont{Pietzsch}},
  \bibinfo{author}{\bibfnamefont{S.}~\bibnamefont{Bl\"{u}gel}},
  \bibnamefont{and}
  \bibinfo{author}{\bibfnamefont{R.}~\bibnamefont{Wiesendanger}},
  \bibinfo{journal}{Nature} \textbf{\bibinfo{volume}{447}},
  \bibinfo{pages}{190} (\bibinfo{year}{2007}).

\bibitem[{\citenamefont{Chen et~al.}(2013)\citenamefont{Chen, Ma, N'Diaye,
  Kwon, Won, Wu, and Schmid}}]{chen2013tailoring}
\bibinfo{author}{\bibfnamefont{G.}~\bibnamefont{Chen}},
  \bibinfo{author}{\bibfnamefont{T.}~\bibnamefont{Ma}},
  \bibinfo{author}{\bibfnamefont{A.~T.} \bibnamefont{N'Diaye}},
  \bibinfo{author}{\bibfnamefont{H.}~\bibnamefont{Kwon}},
  \bibinfo{author}{\bibfnamefont{C.}~\bibnamefont{Won}},
  \bibinfo{author}{\bibfnamefont{Y.}~\bibnamefont{Wu}}, \bibnamefont{and}
  \bibinfo{author}{\bibfnamefont{A.~K.} \bibnamefont{Schmid}},
  \bibinfo{journal}{Nature Comm.} \textbf{\bibinfo{volume}{4}}
  (\bibinfo{year}{2013}).

\bibitem[{\citenamefont{R{\"o}{\ss}ler
  et~al.}(2006)\citenamefont{R{\"o}{\ss}ler, Bogdanov, and
  Pfleiderer}}]{rossler2006}
\bibinfo{author}{\bibfnamefont{U.~K.} \bibnamefont{R{\"o}{\ss}ler}},
  \bibinfo{author}{\bibfnamefont{A.~N.} \bibnamefont{Bogdanov}},
  \bibnamefont{and}
  \bibinfo{author}{\bibfnamefont{C.}~\bibnamefont{Pfleiderer}},
  \bibinfo{journal}{Nature} \textbf{\bibinfo{volume}{442}},
  \bibinfo{pages}{797} (\bibinfo{year}{2006}).

\bibitem[{\citenamefont{Nagaosa and Tokura}(2013)}]{nagaosa2013topological}
\bibinfo{author}{\bibfnamefont{N.}~\bibnamefont{Nagaosa}} \bibnamefont{and}
  \bibinfo{author}{\bibfnamefont{Y.}~\bibnamefont{Tokura}},
  \bibinfo{journal}{Nature Nano.} \textbf{\bibinfo{volume}{8}},
  \bibinfo{pages}{899} (\bibinfo{year}{2013}).

\bibitem[{\citenamefont{Heinze et~al.}(2011)\citenamefont{Heinze, von Bergmann,
  Menzel, Brede, Kubetzka, Wiesendanger, Bihlmayer, and
  Bl{\"u}gel}}]{heinze2011spontaneous}
\bibinfo{author}{\bibfnamefont{S.}~\bibnamefont{Heinze}},
  \bibinfo{author}{\bibfnamefont{K.}~\bibnamefont{von Bergmann}},
  \bibinfo{author}{\bibfnamefont{M.}~\bibnamefont{Menzel}},
  \bibinfo{author}{\bibfnamefont{J.}~\bibnamefont{Brede}},
  \bibinfo{author}{\bibfnamefont{A.}~\bibnamefont{Kubetzka}},
  \bibinfo{author}{\bibfnamefont{R.}~\bibnamefont{Wiesendanger}},
  \bibinfo{author}{\bibfnamefont{G.}~\bibnamefont{Bihlmayer}},
  \bibnamefont{and}
  \bibinfo{author}{\bibfnamefont{S.}~\bibnamefont{Bl{\"u}gel}},
  \bibinfo{journal}{Nature Phys.} \textbf{\bibinfo{volume}{7}},
  \bibinfo{pages}{713} (\bibinfo{year}{2011}).

\bibitem[{\citenamefont{Sampaio et~al.}(2013)\citenamefont{Sampaio, Cros,
  Rohart, Thiaville, and Fert}}]{sampaio2013nucleation}
\bibinfo{author}{\bibfnamefont{J.}~\bibnamefont{Sampaio}},
  \bibinfo{author}{\bibfnamefont{V.}~\bibnamefont{Cros}},
  \bibinfo{author}{\bibfnamefont{S.}~\bibnamefont{Rohart}},
  \bibinfo{author}{\bibfnamefont{A.}~\bibnamefont{Thiaville}},
  \bibnamefont{and} \bibinfo{author}{\bibfnamefont{A.}~\bibnamefont{Fert}},
  \bibinfo{journal}{Nature Nano.} \textbf{\bibinfo{volume}{8}},
  \bibinfo{pages}{839} (\bibinfo{year}{2013}).

\bibitem[{\citenamefont{Iwasaki et~al.}(2013)\citenamefont{Iwasaki, Mochizuki,
  and Nagaosa}}]{iwasaki2013current}
\bibinfo{author}{\bibfnamefont{J.}~\bibnamefont{Iwasaki}},
  \bibinfo{author}{\bibfnamefont{M.}~\bibnamefont{Mochizuki}},
  \bibnamefont{and} \bibinfo{author}{\bibfnamefont{N.}~\bibnamefont{Nagaosa}},
  \bibinfo{journal}{Nature Nano.} \textbf{\bibinfo{volume}{8}},
  \bibinfo{pages}{742} (\bibinfo{year}{2013}).

\bibitem[{\citenamefont{Garello et~al.}(2013)\citenamefont{Garello, Miron,
  Avci, Freimuth, Mokrousov, Bl\"{u}gel, Auffret, Boulle, Gaudin, and
  Gambardella}}]{garello2013}
\bibinfo{author}{\bibfnamefont{K.}~\bibnamefont{Garello}},
  \bibinfo{author}{\bibfnamefont{I.~M.} \bibnamefont{Miron}},
  \bibinfo{author}{\bibfnamefont{C.~O.} \bibnamefont{Avci}},
  \bibinfo{author}{\bibfnamefont{F.}~\bibnamefont{Freimuth}},
  \bibinfo{author}{\bibfnamefont{Y.}~\bibnamefont{Mokrousov}},
  \bibinfo{author}{\bibfnamefont{S.}~\bibnamefont{Bl\"{u}gel}},
  \bibinfo{author}{\bibfnamefont{S.}~\bibnamefont{Auffret}},
  \bibinfo{author}{\bibfnamefont{O.}~\bibnamefont{Boulle}},
  \bibinfo{author}{\bibfnamefont{G.}~\bibnamefont{Gaudin}}, \bibnamefont{and}
  \bibinfo{author}{\bibfnamefont{P.}~\bibnamefont{Gambardella}},
  \bibinfo{journal}{Nature Nano.} \textbf{\bibinfo{volume}{8}},
  \bibinfo{pages}{587} (\bibinfo{year}{2013}).

\bibitem[{\citenamefont{Je et~al.}(2013)\citenamefont{Je, Kim, Yoo, Min, Lee,
  and Choe}}]{Choe}
\bibinfo{author}{\bibfnamefont{S.-G.} \bibnamefont{Je}},
  \bibinfo{author}{\bibfnamefont{D.-H.} \bibnamefont{Kim}},
  \bibinfo{author}{\bibfnamefont{S.-C.} \bibnamefont{Yoo}},
  \bibinfo{author}{\bibfnamefont{B.-C.} \bibnamefont{Min}},
  \bibinfo{author}{\bibfnamefont{K.-J.} \bibnamefont{Lee}}, \bibnamefont{and}
  \bibinfo{author}{\bibfnamefont{S.-B.} \bibnamefont{Choe}},
  \bibinfo{journal}{Phys. Rev. B} \textbf{\bibinfo{volume}{88}},
  \bibinfo{pages}{214401} (\bibinfo{year}{2013}).

\bibitem[{\citenamefont{Metaxas et~al.}(2007)\citenamefont{Metaxas, Jamet,
  Mougin, Cormier, Ferr{\'e}, Baltz, Rodmacq, Dieny, and Stamps}}]{metaxas}
\bibinfo{author}{\bibfnamefont{P.~J.} \bibnamefont{Metaxas}},
  \bibinfo{author}{\bibfnamefont{J.~P.} \bibnamefont{Jamet}},
  \bibinfo{author}{\bibfnamefont{A.}~\bibnamefont{Mougin}},
  \bibinfo{author}{\bibfnamefont{M.}~\bibnamefont{Cormier}},
  \bibinfo{author}{\bibfnamefont{J.}~\bibnamefont{Ferr{\'e}}},
  \bibinfo{author}{\bibfnamefont{V.}~\bibnamefont{Baltz}},
  \bibinfo{author}{\bibfnamefont{B.}~\bibnamefont{Rodmacq}},
  \bibinfo{author}{\bibfnamefont{B.}~\bibnamefont{Dieny}}, \bibnamefont{and}
  \bibinfo{author}{\bibfnamefont{R.~L.} \bibnamefont{Stamps}},
  \bibinfo{journal}{Phys. Rev. Lett.} \textbf{\bibinfo{volume}{99}},
  \bibinfo{pages}{217208} (\bibinfo{year}{2007}).

\bibitem[{\citenamefont{Lacour et~al.}(2007)\citenamefont{Lacour, Hehn, Alnot,
  Montaigne, Greullet, Lengaigne, Lenoble, Robert, and Schuhl}}]{Lacour}
\bibinfo{author}{\bibfnamefont{D.}~\bibnamefont{Lacour}},
  \bibinfo{author}{\bibfnamefont{M.}~\bibnamefont{Hehn}},
  \bibinfo{author}{\bibfnamefont{M.}~\bibnamefont{Alnot}},
  \bibinfo{author}{\bibfnamefont{F.}~\bibnamefont{Montaigne}},
  \bibinfo{author}{\bibfnamefont{F.}~\bibnamefont{Greullet}},
  \bibinfo{author}{\bibfnamefont{G.}~\bibnamefont{Lengaigne}},
  \bibinfo{author}{\bibfnamefont{O.}~\bibnamefont{Lenoble}},
  \bibinfo{author}{\bibfnamefont{S.}~\bibnamefont{Robert}}, \bibnamefont{and}
  \bibinfo{author}{\bibfnamefont{A.}~\bibnamefont{Schuhl}},
  \bibinfo{journal}{Appl. Phys. Lett.} \textbf{\bibinfo{volume}{90}},
  \bibinfo{eid}{192506} (\bibinfo{year}{2007}).

\bibitem[{\citenamefont{Tarasenko et~al.}(1998)\citenamefont{Tarasenko,
  Stankiewicz, Tarasenko, and Ferr{\'e}}}]{Tarasenko}
\bibinfo{author}{\bibfnamefont{S.~V.} \bibnamefont{Tarasenko}},
  \bibinfo{author}{\bibfnamefont{A.}~\bibnamefont{Stankiewicz}},
  \bibinfo{author}{\bibfnamefont{V.~V.} \bibnamefont{Tarasenko}},
  \bibnamefont{and}
  \bibinfo{author}{\bibfnamefont{J.}~\bibnamefont{Ferr{\'e}}},
  \bibinfo{journal}{J. Magn. Magn. Mat.} \textbf{\bibinfo{volume}{189}},
  \bibinfo{pages}{19} (\bibinfo{year}{1998}).

\bibitem[{\citenamefont{Heide et~al.}(2008)\citenamefont{Heide, Bihlmayer, and
  Bl{\"u}gel}}]{heide2008dzyaloshinskii}
\bibinfo{author}{\bibfnamefont{M.}~\bibnamefont{Heide}},
  \bibinfo{author}{\bibfnamefont{G.}~\bibnamefont{Bihlmayer}},
  \bibnamefont{and}
  \bibinfo{author}{\bibfnamefont{S.}~\bibnamefont{Bl{\"u}gel}},
  \bibinfo{journal}{Phys. Rev. B} \textbf{\bibinfo{volume}{78}},
  \bibinfo{pages}{140403} (\bibinfo{year}{2008}).

\bibitem[{\citenamefont{Mihai et~al.}(2013)\citenamefont{Mihai, Whiteside,
  Canwell, Marrows, Benitez, McGrouther, McVitie, McFadzean, and
  Moore}}]{Mihai}
\bibinfo{author}{\bibfnamefont{A.~P.} \bibnamefont{Mihai}},
  \bibinfo{author}{\bibfnamefont{A.~L.} \bibnamefont{Whiteside}},
  \bibinfo{author}{\bibfnamefont{E.~J.} \bibnamefont{Canwell}},
  \bibinfo{author}{\bibfnamefont{C.~H.} \bibnamefont{Marrows}},
  \bibinfo{author}{\bibfnamefont{M.~J.} \bibnamefont{Benitez}},
  \bibinfo{author}{\bibfnamefont{D.}~\bibnamefont{McGrouther}},
  \bibinfo{author}{\bibfnamefont{S.}~\bibnamefont{McVitie}},
  \bibinfo{author}{\bibfnamefont{S.}~\bibnamefont{McFadzean}},
  \bibnamefont{and} \bibinfo{author}{\bibfnamefont{T.~A.} \bibnamefont{Moore}},
  \bibinfo{journal}{Appl. Phys. Lett.} \textbf{\bibinfo{volume}{103}},
  \bibinfo{eid}{262401} (\bibinfo{year}{2013}).

\bibitem[{\citenamefont{Sch\"{u}tz and Fischer}(1992)}]{Fisher_XMCD}
\bibinfo{author}{\bibfnamefont{G.}~\bibnamefont{Sch\"{u}tz}} \bibnamefont{and}
  \bibinfo{author}{\bibfnamefont{P.}~\bibnamefont{Fischer}},
  \bibinfo{journal}{Z. Phys. A Hadrons and Nuclei}
  \textbf{\bibinfo{volume}{341}}, \bibinfo{pages}{227} (\bibinfo{year}{1992}).

\end{thebibliography}

\end{document}